\documentclass[aps,prd,preprintnumbers,superscriptaddress,nofootinbib,floatfix,twocolumn,notitlepage]{revtex4-2}
\usepackage{graphicx}  
\usepackage{dcolumn}   
\usepackage{bm}        
\usepackage{epsfig,amsmath,amssymb,verbatim,mathrsfs,array,layout,textcomp,amssymb,latexsym,slashed}
\usepackage{xcolor}
\usepackage{siunitx}
\usepackage[colorlinks=true,citecolor=blue,urlcolor=blue,linktocpage=true,
linkcolor=blue]{hyperref}
\usepackage[utf8]{inputenc}
\usepackage{multirow}
\usepackage[capitalize]{cleveref}

\usepackage[normalem]{ulem}

\newcommand{\cO}{\mathcal{O}}

\newcommand{\keV}{\mathrm{keV}}

\newcommand{\eg}{\textit{e.g.}}

\def\lag{\mathscr{L}}
\def\bs{\boldsymbol}

\def\beq{\begin{equation}}
\def\eeq{\end{equation}}
\def\beqa{\begin{eqnarray}}
\def\eeqa{\end{eqnarray}}

\newcommand{\bolds}[1]{\boldsymbol{#1}}

\newcommand{\Mpl}{M_{\rm Pl}}

\begin{document}

\title{ 
 Probing New Forces with Nuclear Clocks}
\preprint{LAPTH-007/25}

%
\author{C\'edric Delaunay}
\email{cedric.delaunay@lapth.cnrs.fr}
\affiliation{Laboratoire d'Annecy de Physique Th\'eorique,  74940 Annecy, France}
\author{Seung J. Lee}
\email{sjjlee@kias.re.kr}
\affiliation{School of Physics, KIAS, Seoul 02455, Korea}
\author{Roee Ozeri}
\email{Roee.Ozeri@weizmann.ac.il}
\affiliation{Department of Complex Systems,
Weizmann Institute of Science, Rehovot, Israel 7610001}
\author{Gilad Perez}
\email{gilad.perez@weizmann.ac.il}
\affiliation{Department of Particle Physics and Astrophysics,
Weizmann Institute of Science, Rehovot, Israel 7610001}
\author{Wolfram Ratzinger}
\email{wolfram.ratzinger@weizmann.ac.il}
\affiliation{Department of Particle Physics and Astrophysics,
Weizmann Institute of Science, Rehovot, Israel 7610001}
\author{Bingrong Yu}
\email{bingrong.yu@cornell.edu}
\affiliation{Department of Physics, LEPP, Cornell University, Ithaca, NY 14853, USA}
\affiliation{School of Physics, KIAS, Seoul 02455, Korea}


\begin{abstract}
Clocks based on nuclear isomer transitions promise exceptional stability and precision. The low transition energy of the thorium-229 isomer makes it an ideal candidate, as it  has been excited by a vacuum-ultraviolet laser and is highly sensitive to subtle interactions. This enables the development of powerful tools for probing new forces, which we call {\it quintessometers}. In this work, we demonstrate the potential of nuclear clocks, particularly solid-state variants, to surpass existing limits on scalar field couplings, exceeding the sensitivity of current fifth-force searches at submicron distances and significantly improving equivalence-principle tests at kilometer scales and beyond. Additionally, we highlight the capability of transportable nuclear clocks to detect scalar interactions at distances beyond $10\,$km, complementing space-based missions. 
\end{abstract}

\maketitle

%
\section{Introduction}
%

Nuclear clocks represent a revolutionary advancement in timekeeping, offering precision and stability that surpass current state-of-the-art clocks~\cite{Thirolf:2024xlx}. Unlike atomic clocks, which rely on electronic transitions, nuclear clocks are based on nuclear isomer transitions, making them significantly less affected by environmental factors~\cite{Seiferle:2019fbe,Campbell:2012zzb}. Owing to their inherent robustness, solid-state nuclear clocks, where the transitioning nucleus is embedded in a vacuum ultraviolet (VUV) transparent host crystal, may be able to reach similar stability as their single ion counterparts~\cite{Kazakov_2012}. The low-lying \SI{8.3}{eV} isomer transition of the thorium-229 ($^{229}$Th) nucleus has been identified as a prime candidate for the realization of such clocks~\cite{Helmer:1994zz,Matinyan:1997ih}. Over the last years the knowledge of the transition energy has improved drastically~\cite{Seiferle:2019fbe,Sikorsky:2020peq,Kraemer:2022gpi}, culminating in the first laser excitation at the beginning of last year~\cite{Tiedau:2024obk} (see also~\cite{Elwell:2024qyh}). 
Since then, the error on the frequency was drastically decreased with the use of a frequency comb~\cite{Zhang:2024ngu}. At this rapid rate of technological advancement, the realization of the first nuclear clock is expected very shortly.

Furthermore, the $^{229}$Th isomer transition is exceptionally sensitive to physics beyond the Standard Model (SM). Its remarkably low transition energy most likely arises from the fact that both the ground and excited state are the result of large, nearly
balanced nuclear and electromagnetic contributions, making the transition energy
highly sensitive to small fractional changes~\cite{Flambaum:2006ak,Fadeev:2020bgt,Minkov:2021ovq,Fadeev:2021odr,Caputo:2024doz,Beeks:2024xnc,Minkov:2024wna}. 
The precise degree of cancellation remains unknown and requires a deeper understanding of the structure of $^{229}$Th~\cite{Caputo:2024doz}.
This exceptional sensitivity, combined with the high stability of nuclear clocks, makes them powerful sensors for detecting variations in fundamental constants. 
Such variations could arise from the oscillations of a scalar field in models of ultralight dark matter (ULDM), with nuclear clocks playing a central role for scalar ULDM searches~\cite{Arvanitaki:2014faa,Stadnik:2015kia,Safronova:2017xyt,Fuchs:2024xvc}. 
Beyond this, nuclear clocks could also probe QCD-like axion models~\cite{Kim:2022ype}, detect dark matter (DM) signatures from solar or Earth  subhalos~\cite{Banerjee:2019epw,Banerjee:2019xuy,Budker:2023sex}, and possibly observe
transient events linked to topological structures~\cite{Derevianko:2013oaa}.

In this work, we focus on a minimal scenario where a scalar field is coupled to ordinary matter, without constituting DM or varying due to topological structures. The coupling to matter causes the field to vary near massive objects. These variations could be detected as equivalence-principle (EP) violating effects between atomic and  nuclear clocks, arising from the same interaction~\cite{Damour:1999ip,Damour:2010rp}. (For cases where these bounds are weakened, see~\cite{Oswald:2021vtc, Banerjee:2022sqg}.) Moreover, the scalar field, potentially associated with quintessence~\cite{aristotle_heavens}, induces a fifth-force that can be probed experimentally, placing strong bounds on such scenarios~\cite{Spero:1980zz,Mostepanenko:1990ed,Bordag:1993htq,Lamoreaux:1996wh,Bordag:1998nv,Ederth:2000zz,Harris:2000zz,Fischbach:2001ry,Chiaverini:2002cb,Adelberger:2003zx,Long:2003dx,Kapner:2006si,Yang:2012zzb,Chen:2014oda,Leefer:2016xfu,Tan:2020vpf,Lee:2020zjt}.  Here, we argue that nuclear clocks serve as highly precise sensors of this type of interaction, which we therefore refer to as {\it quintessometers}. Note that this is a new term introduced in this work, and the scalar field here is not necessarily related to any cosmological model of dark energy.
We identify unconstrained regions of parameter space that can be probed using nuclear clocks, building on the work of~\cite{Brzeminski:2022sde} that focused on spaceborne missions. We emphasize the potential of transportable clocks to constrain scalar interactions with ranges greater than $\mathcal{O}(\SI{10}{\km})$, even before reaching space. Furthermore, we demonstrate that nuclear clocks can 
surpass existing fifth-force searches at distances smaller than $\mathcal{O}(\SI{100}{\nm})\,$. These proposals take advantage of solid-state nuclear clocks, where the nucleus is confined to scales of $\mathcal{O}(0.1-\SI{1}{\nm})\,$, making such measurements impossible using single-ion clocks with comparable stability.

In~\cref{sec:formalism}, we lay out the formalism used in our analysis. In~\cref{sec:field solution from a macroscopic probe}, we solve the classical equation of motion for a linearly coupled scalar in the presence of a macroscropic source. In~\cref{sec:experimental concepts}, we explore general strategies for probing a light scalar field using nuclear clocks. We then examine specific detection scenarios at both large and small distances in~\cref{sec:large-distance}  and~\cref{sec:short-distance}, respectively. Using nuclear clocks, we derive prospective bounds on the scalar coupling across previously unconstrained regions of parameter space for various length scales, which are shown in~\cref{fig:moneyplots}. Finally, we present our conclusions in~\cref{sec:conclusion}.

%
\section{Linearly coupled scalar}
\label{sec:formalism}
%
Consider a scalar field $\phi$ interacting with SM fields through CP-even linear couplings. Its interaction Lagrangian is given by,
 \begin{align}
 \label{eq:Lint}
    \lag_{\rm int} &= \bigg[
    \frac{d_e}{16\pi\alpha}F_{\mu\nu}F^{\mu\nu}
     - \frac{d_g\beta(\alpha_s)}{4\alpha_s}G_{\mu\nu}^a G^{a\mu\nu}
    - d_{m_e}m_e\, \bar{e}e \nonumber
    \\&\qquad
    - \sum_{q=u,d,s}\left(d_{m_q} 
    + \gamma_{m_q}d_g\right) m_q\bar{q} q
    \bigg] \frac{\phi}{\Mpl}\,,
 \end{align}
where $G_{\mu\nu}^a$  and $F_{\mu\nu}$ are the gluon and the photon field-strengths, respectively, $\alpha$ is the fine structure constant, $\alpha_s$ denotes the QCD coupling, and $\beta(\alpha_s)=-(11-2n_f/3)\alpha_s^2/2\pi$ is its one-loop $\beta$-function, with  $n_f=3$ light quark flavors. The  electron, along with the up, down and strange quark fields, are denoted by $e$ and $q=u,d,s$, with corresponding masses $m_e$ and $m_q$, respectively. The $\gamma_{m_q}$'s represent the anomalous dimensions of the quark masses due to QCD interactions. 

The $d$'s are dimensionless constants that quantify the coupling strengths of $\phi$ to SM fields relative to gravitational interactions, characterized by the reduced Planck mass $\Mpl\approx\SI{2.4e18}{GeV}$. The other factors are chosen so that these constants directly express the dependence of $\alpha$, the QCD confinement scale $\Lambda_{\rm QCD}$, and the fermion masses on the dimensionless parameter $\varphi\equiv\phi/\Mpl$, {\it i.e.},
\beq
    \frac{{\rm d}\log\alpha}{{\rm d}\varphi}=d_e\,,\ 
    \frac{{\rm d}\log\Lambda_{\rm QCD}}{{\rm d}\varphi}=d_g\,,\ 
    \frac{{\rm d}\log m_{e,q}}{{\rm d}\varphi}=d_{m_{e,q}}\,,\label{eq:convention1}
\eeq
where the $m_q$'s are evaluated at $\Lambda_{\rm QCD}$.

Another commonly used convention for parameterizing the effect of an ultralight scalar on physical observables is given by,
\begin{align}
    \label{eq:convention2}
    \hat{O}_i (\phi) = \hat{O}_i \left(1+\frac{\phi}{\Lambda_i}\right),
\end{align}
where $i=e,g,m_e,m_q$ correspond to the observables $\hat{O}_i=\alpha,\Lambda_{\rm QCD},m_e,m_q$, respectively, and the $\Lambda_i$'s represent the new-physics scales associated with the coupling of $\phi$ to SM fields. For example, the photon coupling in~\cref{eq:convention2} takes the form $\phi F^{\mu\nu}F_{\mu\nu}/(4 e^2 \Lambda_e)$. Throughout this paper, we adopt the normalization in~\cref{eq:Lint}, but the relationship between $d_i$ and $\Lambda_i$ remains straightforward,
\begin{align}
\Lambda_i = \frac{M_{\rm Pl}}{d_i} = 2.4\,{\rm TeV}\, \left(\frac{10^{15}}{d_i}\right).       
\end{align}

From~\cref{eq:convention1}, the dependence of an atomic mass $m_A$ on the field $\varphi$ 
can be expressed as,
\beq\label{eq:definition dilatonic charges sensitivities}
    \frac{{\rm d}\log m_A}{{\rm d}\varphi}=\sum_i Q_i d_i\equiv Q\cdot d\,, 
\eeq
where the sum runs over $i=e,g,m_e,m_u,m_d,m_s$, and $Q_e\equiv {\rm d}\log m_A/{\rm d}\log \alpha$, etc.
The coefficients $Q_i$, known as dilatonic charges, depend on the specific atom under consideration (see, \eg~\cite{Damour:2010rp}). A universal approximation of $Q_g\approx 1$ is often valid, as the QCD binding energy overwhelmingly dominates atomic masses. Similarly, atomic and nuclear transition frequencies $\nu$ follow an analogous decomposition, 
\beq\label{eq:definition sensitivity coefficients}
\frac{{\rm d}\log \nu}{{\rm d}\varphi}=\sum_i K_i d_i\equiv K\cdot d\,,
\eeq
where the transition-dependent parameters $K_i$ are commonly referred to as sensitivity coefficients. 

The coefficients $K_g$ and $K_e$ can be roughly estimated as follows. The energy difference $\Delta E\sim \mathcal{O}(10\,$eV) between the ground and isomeric states of $^{229}$Th is significantly smaller than the typical MeV-scale values observed in all other nuclear excitations. This is likely due to an accidental cancellation between the strong force and electromagnetic contributions, $\Delta E_{\rm nuc}$ and $\Delta E_{\rm em}$, respectively, to the nuclear binding energy. As a result, any variation in the strong or electromagnetic contributions is expected to be largely amplified in the transition frequency $\nu$, 
\beq
\frac{\delta \nu}{\nu}\sim K_g \frac{\delta(\Delta E_{\rm nuc})}{\Delta E_{\rm nuc}}+K_e\frac{\delta(\Delta E_{\rm em})}{\Delta E_{\rm em}}\,, 
\eeq
with 
$K_{g,e}\equiv \Delta E_{\rm nuc,em}/\Delta E\sim \mathcal{O}(10^5)$. More detailed modeling of the thorium nucleus supports this estimate~\cite{Fadeev:2020bgt,Fadeev:2021odr,Caputo:2024doz}. Although a scenario in which $\Delta E_{\rm nuc,em}\ll\,$MeV leads to weaker sensitivity to new physics remains a possibility, it is considered unlikely~\cite{Caputo:2024doz}.

%
\section{Field solution from a macroscopic probe}
\label{sec:field solution from a macroscopic probe}
%
Given the interaction Lagrangian in~\cref{eq:Lint}, a static macroscopic probe generates a nonzero $\phi$ field, which satisfies the time-independent Klein-Gordon equation,
\beq
\left(\bs{\nabla}^2 -m_\phi^2\right)\phi(\bolds{r})=\frac{Q(\bolds{r})\cdot d}{\Mpl} \rho(\bolds{r})\,,
\eeq
where $\rho(\bolds{r})$ is the probe's mass density, $m_\phi$ is the mass of $\phi$, and bold symbols denote spatial vectors. The general solution is, 
\beq\label{eq:solphi}
\phi(\bolds{r})=\frac{1}{\Mpl} \int {\rm d}^3\bs{r'}\, \phi_G(\bolds{r},\bs{r'})\  Q(\bolds{r}')\cdot d\  \rho(\bolds{r'})\,,
\eeq
where $\phi_G(\bolds{r},\bolds{r'})\equiv -e^{-m_\phi|\bolds{r}-\bolds{r'}|}/(4\pi|\bolds{r}-\bolds{r'}|)$ is Green's function. 
For a homogeneous spherical probe of radius $R$ and mass $M$, \cref{eq:solphi} simplifies outside the sphere to
\beq\label{eq:solphi-sphere}
\phi\left(r>R\right)=- Q\cdot d  \frac{M}{\Mpl} \frac{e^{-m_\phi r}}{4\pi r} F(m_\phi R)\,,
\eeq
where $F(x)\equiv 3\left(x\cosh x-\sinh x\right)/x^3$. If the probe consists of multiple atomic species, the $Q$ values are mass-fraction-weighted averages. 
Inside the sphere, the field solution is,
\beq\begin{aligned}
\phi(r\leq R) &= Q\cdot d \frac{M}{\Mpl}\left[\frac{\sinh(m_\phi r)}{4\pi r}f(m_\phi R)\right.\\
&\hspace{7em}
\left.-(m_\phi R)^{-2}\frac{3}{4\pi R}\right]\,,
\end{aligned}
\eeq
where $f(x)\equiv 3e^{-x}(1+x)/x^3$. 

There are two limiting behaviors of interest. First, the form factor $F(m_\phi R)$ asymptotes to unity as $m_\phi R\to 0$, corresponding to a point-like source, with,
\beq\label{eq:phi-pointlike}
\phi_{\rm point}(r)\approx -Q\cdot d\frac{M}{M_{\rm Pl}}\frac{e^{-m_\phi r}}{4\pi r}\,.
\eeq
On the other hand, in the large sphere limit $m_\phi R\gg1$, in the vicinity of the sphere's surface at a distance $r=R+ s$ with separation $|s|\ll R$, the field solution approximates that of an infinite plate, 
\begin{align}
    \phi_{{\rm plate}}(R+s) &\approx
    - Q\cdot  d\,\frac{ \  \rho }{2 M_{\rm Pl}m_\phi^2}\,\left\{\begin{array}{cc} e^{-m_\phi s} &\,, s>0\\ 2 - e^{-m_\phi |s|} &\,, s<0\end{array}\right.\,,\label{eq:phi_one_sided_infinite_medium}
\end{align}
where $\rho\equiv 3M/(4\pi R^3)$ is the probe's density, which no longer depends on $R$. Note that in both cases, the field is exponentially suppressed at distances larger than $\sim 1/m_\phi$ from the source. In the $s/R\rightarrow0$ limit,~\cref{eq:phi_one_sided_infinite_medium} can be interpreted as only the material within a distance $\sim 1/m_\phi$ sourcing the field. Therefore, any object extending much further than $1/m_\phi$ may be approximated by this formula. 

%
\section{Experimental concepts}
\label{sec:experimental concepts}
%

Suppose a probe, as discussed above, is brought near a nuclear clock, and that their relative distance is varied in a controlled manner, causing the background field sourced by the probe to change at the position of the nuclear clock. This variation $\delta\phi$ will induce a change in the fundamental parameters, and thus in the clock frequency, according to~\cref{eq:definition sensitivity coefficients},
\beq
    \frac{\delta\nu_{\rm Th}}{\nu_{\rm Th}}=  K_{\rm Th}\cdot  d\ \frac{\delta \phi}{\Mpl}\,.
\eeq

There are two ways in which an experiment might detect this frequency change~\cite{Brzeminski:2022sde}. The first way is to compare the nuclear clock with a second clock based on an electronic transition, at the same location. In the absence of new physics the frequency ratio $\nu_{\rm Th}/\nu_{\rm el}$ is position- and time-independent. However, in the presence of the field $\phi$, 
the ratio will change as the two clocks respond differently to variations in the fundamental parameters. Hence,
\beq
\frac{\delta(\nu_{\rm Th}/\nu_{\rm el})}{\nu_{\rm Th}/\nu_{\rm el}}=(K_{\rm Th}-  K_{\rm el})\cdot   d \ \frac{\delta \phi}{\Mpl}\,.
\label{eq:difference}
\eeq
Since, generically, $K_{\rm Th}\gg K_{\rm el}$~\cite{Kim:2023pvt}, such comparisons are not expected to suppress the sensitivity of the nuclear clock to new CP-even scalars \footnote{Given the gravitational acceleration on the surface of earth $g\simeq \SI{10}{m/s^2}$, one would have to place the two clocks to within $\sim\SI{1}{mm}$ at the same height in order for the gravitational redshift to be negligible at an accuracy of $10^{-19}$. Potentially significantly smaller offsets may be achieved by using an electronic transition of the same thorium atom.}. 
Alternatively, one might compare the frequencies of two (identical) nuclear clocks at different locations via an optical link. In this case, the frequency change will receive an additional contribution from time dilation due to the difference in gravitational potential $\delta\Psi_{\rm grav}$ between the two positions,
\beq
\frac{\delta\nu_{\rm Th}}{\nu_{\rm Th}}=\delta\Psi_{\rm grav}+K_{\rm Th}\cdot d \ \frac{\delta\phi}{\Mpl}\,.
\label{eq:grav}
\eeq
We note that in this setup other tests of general relativity than the one discussed in this work can be carried out~\cite{Derevianko:2021kye}.  To keep the discussion simple, we will not explore this approach further and only consider the comparison of a nuclear clock to one based on an electronic transition in~\cref{sec:large-distance}. All experiments suggested there could also be conducted using this method with only minor modifications. The experiments discussed in~\cref{sec:short-distance} only reach coupling values $d\gg 1$ making relativistic effects negligible and a distinction between the methods superfluous.

Clearly, the larger the value of $\delta\phi$, the higher the sensitivity. One can change the field value by simply varying the distance between the source and the nuclear clock, {\it i.e.} $\delta\phi=\phi(r_2)-\phi(r_1)$. There are a few special cases of particular interest: 

\begin{itemize}

\item 
In the point-source limit ($m_\phi R\to 0$), by taking $r_2\gg r_1$ to maximize the field variation, \cref{eq:solphi-sphere} gives,
 \beq
 \delta\phi_{\rm point}\propto \frac{e^{-m_\phi r_1}}{r_1}\,,
 \eeq
which vanishes exponentially for $m_\phi\gg r_1^{-1}$, with sensitivity to the $d$ parameters independent of $m_\phi$ in the limit $m_\phi r_1\to 0$.

\item In contrast, in the infinite-plate limit ($m_\phi R\to \infty$), \cref{eq:phi_one_sided_infinite_medium} gives,
\beq
\delta\phi_{{\rm plate}}\propto m_\phi^{-2}\left(e^{-m_\phi s_1}-e^{-m_\phi s_2}\right)\,,
\eeq
where $s_{1,2}\equiv r_{1,2}-R\ll R$. In this case, three distinct regimes can be identified. For $m_\phi \gg s_{1,2}^{-1}$, $\delta\phi$ vanishes exponentially. For $m_\phi\ll s_{1,2}^{-1}$, $\delta\phi\propto m_\phi^{-1}$, leading to a sensitivity on $d\propto\sqrt{m_\phi}$. In the intermediate regime where $s_1^{-1}\ll m_\phi\ll s_2^{-1}$, $\delta\phi\propto m_\phi^{-2}$, and the sensitivity weakens linearly with increasing $m_\phi$.

\end{itemize}

Recent laser excitation of the $^{229}$Th nucleus isomeric transition~\cite{Tiedau:2024obk,Elwell:2024qyh} has significantly improved the accuracy of its frequency, reaching $\delta\log\nu_{\rm Th}\sim \cO(10^{-12})$~\cite{Zhang:2024ngu}. This breakthrough enhances the expectation that, in the near future, the isomeric transition could be monitored by a nuclear clock as stable as current electronic counterparts, {\it i.e.} with $\delta\log\nu=\mathcal{O}(10^{-19})$~\cite{Zhang:2024ngu}. 

\begin{figure}[t]
    \centering
        \includegraphics[width=0.5\textwidth]{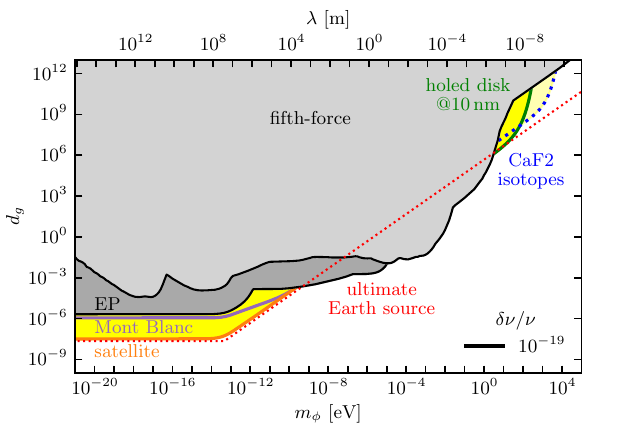}
      \\
        \includegraphics[width=0.5\textwidth]{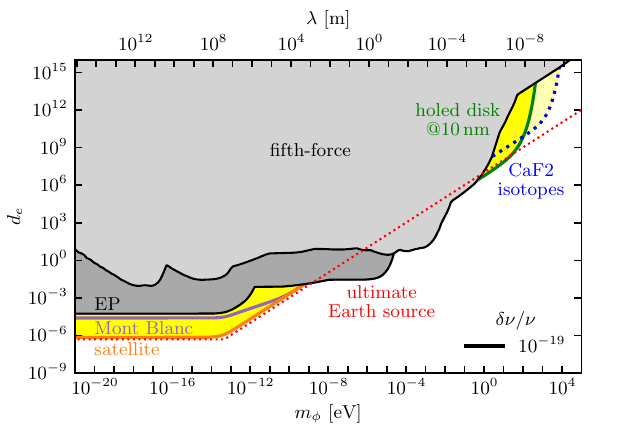}
    \caption{Projected sensitivities of nuclear clock-based quintessometers on the scalar couplings $d_{g}$ (top) and $d_e$ (bottom). The gray-shaded regions represent the parameter space excluded by fifth-force experiments~\cite{Spero:1980zz,Mostepanenko:1990ed,Bordag:1993htq,Lamoreaux:1996wh,Bordag:1998nv,Ederth:2000zz,Harris:2000zz,Fischbach:2001ry,Chiaverini:2002cb,Adelberger:2003zx,Long:2003dx,Kapner:2006si,Yang:2012zzb,Chen:2014oda,Tan:2020vpf,Lee:2020zjt} (lighter gray) and EP tests~\cite{Schlamminger:2007ht,Smith:1999cr,Berge:2017ovy} (darker gray). The yellow-shaded regions indicate the unexplored parameter space that could be explored by quintessometers with a sensitivity of $\delta \nu/\nu\sim 10^{-19}$, and assuming $K_{g,e}= 10^5$.    See~\cref{fig:large_dist} and \cref{fig:small_dist} for zoomed-in views of the large- and short-distance regions, respectively. The red dotted line represents the ultimate quintessometer's sensitivity to scalar fields sourced by the Earth at this frequency uncertainty. The dashed blue line denotes the sensitivity of 
the isotope comparison discussed in~\cref{sec:host crystals that differ by isotope}, achievable once nuclear effects are properly accounted for.}
    \label{fig:moneyplots}
\end{figure}

In the following sections, we estimate the reach of such nuclear clocks to the couplings $d_g$ and $d_e$ of a light scalar to the QCD and QED sectors, respectively, assuming $K_{g,e}\sim \cO(10^5)$. Before delving into the details, it is useful to consider the strongest bound that can be achieved using Earth as the source of the scalar field $\phi$. This bound can be obtained by comparing
the frequency ratio between a nuclear and an electronic clock on Earth's surface with a similar comparison conducted at a far distance. If the range of the force is large compared to Earth's radius ($m_\phi\lesssim R_\oplus^{-1}\approx 3\times 10^{-14}\,$eV), one can impose the constraint,
\beq
|d_g|\approx 3\times 10^{-8} \, \left[\frac{5.5~{\rm g/cm^3}}{\rho_\oplus}\right]^{1/2}\left[\frac{K_g}{10^{-5}}\right]^{-1/2}\left[\frac{\delta\nu/\nu}{10^{-19}}\right]^{1/2}\,,
\label{eq:sensitivity_estimate}
\eeq
where $\rho_\oplus$ is the average density of the Earth.
On the other hand, if the range is shorter ($m_\phi\gtrsim 3\times 10^{-14}\,$eV), Earth behaves like an infinite plate for a clock on its surface, resulting in a limit that is weaker than the one above by a factor $m_\phi/(3\times 10^{-14}\,{\rm eV})$. 
This ultimate reach of a nuclear clock-based quintessometer at fractional sensitivity of $\delta\nu/\nu=10^{-19}$ is represented by the red dotted line in~\cref{fig:moneyplots}.

A comparison with existing bounds from searches for EP violation and fifth-forces (shown in dark and light gray in~\cref{fig:moneyplots}) clearly demonstrates that nuclear clock searches could be competitive for masses $m_\phi\lesssim 10^{-9}\,$eV and $m_\phi\gtrsim 1\,$eV. In the following two sections, we propose experimental setups capable of probing these regimes (yellow shaded regions). 

In the large mass regime, carefully designed source masses will be used instead of the entire Earth. However, since the densities of solids on Earth are at most a few times the average density, the  rough bound derived above remains applicable. Furthermore, when probing the coupling $d_e$, all bounds are weaker by a factor of $\sqrt{Q_e}\sim\mathcal{O}(10^{-1}-10^{-2})$ due to the smaller charge $Q_e$ of the source body. Fifth-force bounds are further suppressed by another factor of $\sqrt{Q_e}$, as the coupling to the test mass is also reduced. Notably, the charge difference $\Delta Q_e$ between two test masses in an EP violation search is not necessarily smaller than $\Delta Q_g$. This suppression enhances the reach of nuclear clock searches compared to conventional fifth-force experiments in the large-mass regime. A similar effect occurs when considering the coupling to quark masses, $d_{m_q}$.

%
\section{Searches at large distances}
\label{sec:large-distance}

\subsection{Space-based nuclear clock}

Consider a satellite mission carrying a nuclear and an atomic clock on board on an elliptic orbit around the Earth, with the maximum and minimum altitudes denoted by $s_{\rm max}$ and $s_{\rm min}\ll s_{\rm max}$, respectively. Assuming $s_{\rm min}$ is comparable to or larger than the Earth's radius $R_\oplus\approx 6400\,$km, it is possible to probe scalars with masses $m_\phi\lesssim s_{\rm min}^{-1}$ using only data from this nuclear-to-atomic clock  comparison, as the satellite experiences different values of the $\phi$ field along its orbit. 
In this mass range, the Earth can be approximated as a point source, and \cref{eq:phi-pointlike} applies. The largest field variation  is given by $\delta\phi\propto s_{\rm min}^{-1} -s_{\rm max}^{-1}\simeq s_{\rm min}^{-1}$, which allows probing coupling values as small as the one given in \cref{eq:sensitivity_estimate} only reduced by a factor $\sqrt{R_\oplus/s_{\rm min}}$\,.
In \cref{fig:large_dist} we show the resulting sensitivity as a thick orange line assuming $s_{\rm min}=2R_\oplus$
In this case, the sensitivity is determined by the stability of the clock-comparison test, as the data collected on board is only compared to itself. As in this setup there is no clock on the ground, the sensitivity drops exponentially for masses $m_\phi>1/R_{\oplus}$ rather than linearly as in the ultimate bound.

With a more ambitious setup, where the frequency ratio is measured with an accuracy comparable to the system's stability, this ultimate bound may be reached. 
To achieve this, any systematic shift not common to both the space-based and ground-based measurements must be eliminated. In this case, the frequency ratio measured onboard can be compared to results obtained on the ground. For masses $m_\phi\gtrsim R_\oplus^{-1}$, the field is exponentially suppressed on the satellite, but ground-based measurements will still be influenced. In this way, the linear scaling is recovered, as shown by the thin orange line in~\cref{fig:moneyplots,fig:large_dist}, which saturates the ultimate reach. With an accuracy of $\delta\nu/\nu\sim 10^{-19}$, this experiment could probe $d_g$ coupling values approximately two orders of magnitude below the current bounds from EP tests~\cite{Berge:2017ovy,MICROSCOPE:2022doy,Smith:1999cr,Schlamminger:2007ht}.  

\subsection{Transportable nuclear clock on the ground}

Since realizing an atomic--or even more so, a nuclear--clock in space presents multiple technological challenges, one may consider a ground-based transportable version, similar to the setup in~\cite{Grotti:2017ftq}, accompanied by another optical clock for comparison. Such a system could be moved around at various altitudes on the ground to probe different values of the field $\phi$ sourced by the Earth. For instance, one might compare the clock frequency ratio at sea level ($s_{\rm min}=0$) with that at a high-altitude mountain ($s_{\rm max}=h$), or alternatively at the bottom of a mine. 
In the point-source regime ($m_\phi\ll R_\oplus^{-1}$), where potential difference is independent of mass, the sensitivity remains constant, and we have, 
\beq
|d_g|\approx 2\times 10^{-6}\,\left[\frac{1\,{\rm km}}{h}\right]^{1/2}\left[\frac{5.5\,{\rm g/cm}^3}{\rho_\oplus}\right]^{1/2}\left[\frac{\delta\nu/\nu}{10^{-19}}\right]^{1/2}\,. 
\eeq
In the infinite-plate limit ($m_\phi\gg R_\oplus^{-1}$), where~\cref{eq:phi_one_sided_infinite_medium} applies, 
the scaling $|d_g|\propto m_\phi$ ($|d_g|\propto \sqrt{m_\phi}$) is recovered for $m_\phi\gg h^{-1}$ ($m_\phi\ll h^{-1}$). 

\begin{figure}[t]
    \centering
        \includegraphics[width=0.5\textwidth]{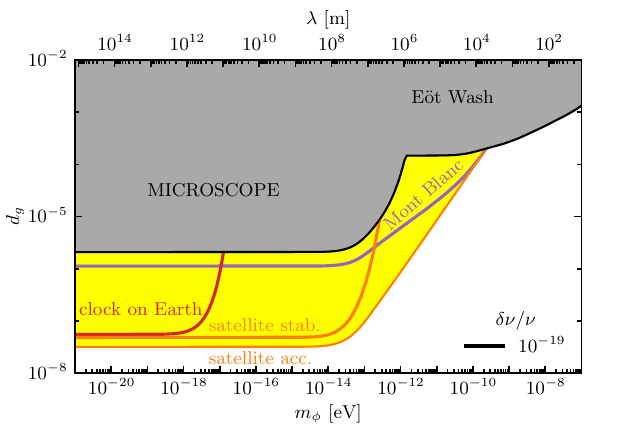}
      \\
        \includegraphics[width=0.5\textwidth]{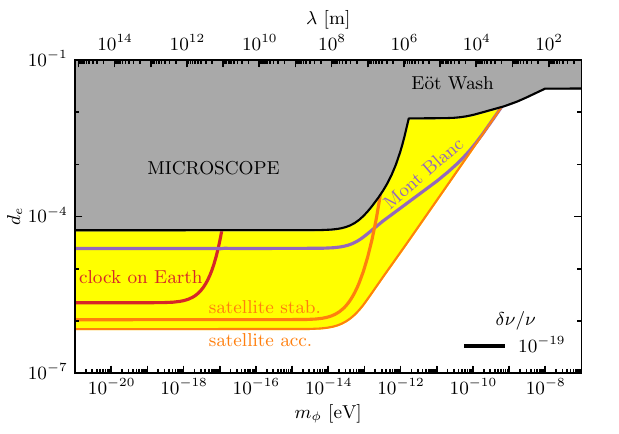}
    \caption{
    Zoomed-in view of the large-distance region of~\cref{fig:moneyplots}, illustrating the reach of the proposed nuclear clock-based quintessometers (yellow) compared to existing bounds from relative acceleration (dark gray). A nuclear clock operating on earth for one year covers the region enclosed by the red line, due to the Earth's eccentric orbit around the Sun. Similarly, a satellite clock with high orbital eccentricity extents the reach to the area enclosed by the thick orange line. If the satellite clock is both stable and accurate to the quoted level, the reach further extends to the thin orange line. An Earth-based transportable experiment, where the height of the clock is varied from sea level to a \SI{4}{\km} altitude, is shown in purple.}
    \label{fig:large_dist}
\end{figure}

For illustration, \cref{fig:moneyplots,fig:large_dist} show the expected sensitivities for $d_{g,e}$ from a ground-based experiment comparing the nuclear-to-optical clock-frequency ratio between sea level and a high-altitude point, such as the Mont Blanc peak. The Vallot observatory on Mont Blanc at $h\approx4\,$km could be a potential location for such an experiment. An accuracy of $\delta\nu/\nu\sim 10^{-19}$ would allow reaching approximately one order of magnitude beyond the current bounds from EP tests for scalar masses around $m_\phi\sim10^{-11}\,$eV.

%
\section{Searches at short distances}
\label{sec:short-distance}
%

%
\subsection{Rotating holed disk}
%

\begin{figure}[t]
    \centering
        \includegraphics[width=0.5\textwidth]{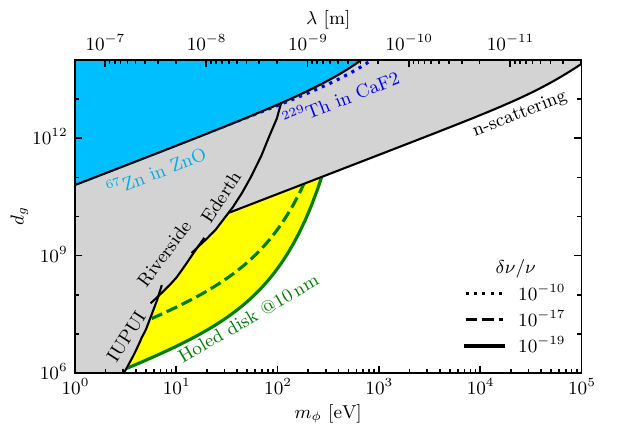}
      \\
        \includegraphics[width=0.5\textwidth]{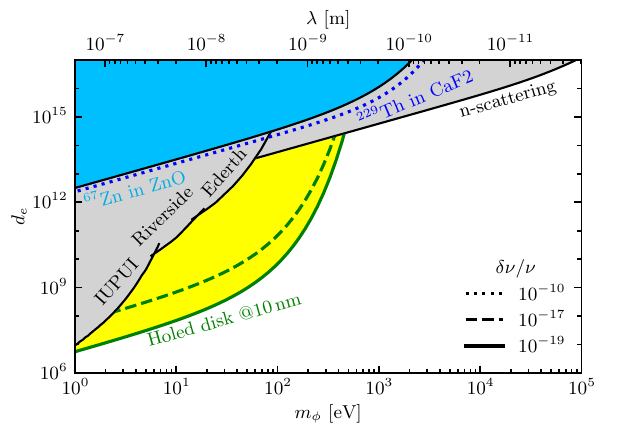}
    \caption{Zoomed-in view of the small-distance region of~\cref{fig:moneyplots}, illustrating the expected sensitivities from placing a rotating disk at a distance of $\SI{10}{\nm}$ from a nuclear clock (green), and from CaF$_2$ crystals made of different Ca isotopes doped with $^{229}$Th. The cyan area represents the bound derived from the non-observation of a M\"ossbauer shift in ZnO due to enrichment (see~\cref{sec:Moessbauer classic}). The light gray region is constrained by fifth-force searches~\cite{Ederth:2000zz,Harris:2000zz,Fischbach:2001ry,Chen:2014oda} and neutron scattering experiments~\cite{Nesvizhevsky:2007by}.  }
    \label{fig:small_dist}
\end{figure}

Consider a large source ($R\gg m_\phi^{-1}$) brought to a short distance $s_1\ll R$ from a nuclear clock, such that the field solution follows~\cref{eq:phi_one_sided_infinite_medium}. The ticking of this clock can be compared to that of an identical clock placed further away ($s_2\to \infty$), thereby maximizing $\delta\phi$.
For scalar masses in the range $R^{-1}\ll m_\phi\ll s_1^{-1}$, the field variation decays $\delta\phi\propto m_\phi^{-2}$, such that the sensitivity to $d$ couplings degrades only linearly with increasing $m_\phi$, while it drops exponentially for $m_\phi\gtrsim s_1^{-1}$. Consequently, reducing the separation $s_1$ extends the mass range over which the sensitivity follows this linear scaling. 
Assuming sources as dense as tungsten, gold or platinum, with $\rho_{\rm W}\approx 19.3\,$g$/$cm$^3$, the sensitivity could reach couplings as small as, 
\beq\label{eq:disksensitivity}
|d_g|\approx 10^7\, \left[\frac{m_\phi}{20\,{\rm eV}}\right]\left[\frac{\rho_{\rm Pt}}{\rho}\right]^{1/2}\left[\frac{\delta\nu/\nu}{10^{-19}}\right]^{1/2}\,,
\eeq
for $m_\phi\lesssim s_1^{-1}\sim 20\,{\rm eV}$ corresponding to $s_1=10\,$nm. A separation of $30\,$nm, along with surface roughness of $\sim1\,$nm, have already been achieved in a similar experimental context~\cite{Chen:2014oda}.  
\cref{fig:moneyplots} and~\cref{fig:small_dist} illustrate the expected sensitivities over a wide range of $m_\phi$ (green lines). For $m_\phi\sim 10\,$eV, the reach could extend about two orders of magnitude beyond current bounds from Casimir force experiments~\cite{Bordag:2001qi}. Although the reach increases for smaller masses and correspondingly larger separations, it remains dominated by existing fifth-force searches for masses smaller than approximately $\sim \SI{1}{eV}$.

This setup could be practically implemented by fabricating a disk of thickness $R\gg m_\phi^{-1}$,   with regularly spaced perforations along a circular path, where the hole diameters and spacing are much larger than $s_1$. Rapidly rotating this disk near the nuclear clock would create a periodic modulation of the frequency, leading to sidebands separated by multiples of the modulation frequency $\nu_{\rm mod}$, corresponding to the rate at which the holes pass by the doped crystal.  

We now outline an interrogation scheme based on sidebands and provide a rough sensitivity estimate. The periodic influence of the dense sections of the rotating disk modulates the isomeric transition frequency in a thin surface layer of the crystal at a frequency $\nu_{\rm mod}$. Consequently, $^{229}$Th nuclei within this layer can be excited by a laser whose frequency is detuned by $\pm \nu_{\rm mod}$ from the unperturbed transition frequency in the absence of the disk. The excitation rate at this detuned frequency is suppressed by a factor of $R_1=|J_{1}(\alpha)|^2R$ compared to the rate $R$ in the absence of modulation, where $J_{1}$ is the order-one Bessel function of the first kind, and $\alpha=\delta\nu/\nu_{\rm mod}$ is the modulation index. To ensure that only these modulated nuclei are excited, the modulation frequency must be much larger than the transition linewidth in the crystal, {\it i.e.}, $\Gamma\ll\nu_{\rm mod}$. For CaF$_2$, the linewidth is dominated by random dipole interactions with neighboring nuclei and is predicted to be  $\Gamma\sim \SI{200}{\Hz}$~\cite{Tiedau:2024obk}. Recent measurements however found linewidths about 2 orders of magnitude larger due to unidentified sources \cite{Ooi:2025hmd}. Below we optimistically assume that these difficulties can be overcome by usage of more advanced crystals.
For small $\alpha$, we use the approximation $|J_{1}(\alpha)|^2=\alpha^2/4$. Since we are interested in small frequency variations, and thus small $\alpha$, we consider the regime where the excitation rate $R_1$ is much smaller than the decay rate in the crystal, $\gamma\simeq 1/\SI{630}{\s}$. After illuminating the crystal for a duration of approximately $1/\gamma$, an equilibrium is reached where a fraction of $\sim R_1/\gamma$ of the nuclei experiencing the scalar field $\phi$ are excited. Assuming the crystal has a cylindrical shape with a front area $A$ aligned with the disk, the number of affected $^{229}$Th nuclei is approximately $\sim n_{\rm Th}A/m_\phi$, under the assumption that $1/m_\phi\gtrsim s_1$, where $n_{\rm Th}$ is the thorium dopant density. The total number of excited nuclei is then given by
\beq
    N_{\rm exc}\sim n_{\rm Th}\frac{A}{m_\phi}\frac{\alpha^2}{4}\frac{R}{\gamma}\,.
\eeq
After the laser is turned off, these excited nuclei emit a signal photon within a time of order $1/\gamma$, which can be detected with probability $p_{\rm sig}$, leading to a total signal count of $N_{\rm sig}=p_{\rm sig}N_{\rm exc}$. To detect the effect, the signal count must be large enough, {\it i.e.}, $N_{\rm sig}\gtrsim\sqrt{N_{\rm bkg}}$, where the background count $N_{\rm bkg}$ is primarily due to radioluminescence from the $\alpha$-decays of $^{229}$Th~\cite{Tiedau:2024obk}, which has a half-life of $\tau_{\alpha}\sim\SI{8000}{yr}$. Conservatively assuming that this background originates from the whole volume of the crystal, the count rate is given by
\beq
    N_{\rm bkg}\sim p_{\rm bkg} n_{\rm Th} \frac{Ad }{\tau_{\alpha}\gamma}\,,
\eeq
where $d$ is the thickness of the crystal and $p_{\rm bkg}$ is the probability that a photon emitted after an $\alpha$-decay is misidentified as a signal photon. Combining these expressions, the  sensitivity to fractional frequency shifts is,
\beq
    \frac{\delta\nu}{\nu}=\frac{2\nu_{\rm mod}}{\nu}\left(\frac{p_{\rm bkg}}{p^2_{\rm sig}}\frac{m_\phi^2 d}{n_{\rm Th} A}\frac{\gamma}{R^2\tau_{\alpha}}\right)^{1/4}\,.
\eeq
For typical experimental parameters, {\it i.e.}, $\nu_{\rm mod}=\SI{1}{\kHz}$, $p_{\rm bkg}=p_{\rm sig}=1\%$~\footnote{These values roughly correspond to the ones realized in the apparatus of  Ref.~\cite{Tiedau:2024obk}, where $p_{\rm sig}=2\times10^{-3}$ was estimated and a crystal with a $66\,{\rm kBq}$ $^{229}$Th activity lead to a $~400\,{\rm s}^{-1}$ background count rate ($p_{\rm bkg}\approx 6\times10^{-3}$).}, $m_\phi=1/\SI{10}{\nm}$, $d=\SI{1}{\micro\metre}$, $A=\SI{1}{\mm^2}$, $n_{\rm Th}=\SI{e19}{\cm^{-3}}$ and $R=10\gamma$ (corresponding to a laser intensity of $I\sim \SI{20}{\milli\W/\cm^2}$~\cite{Kazakov_2012}) we find,
\beq
    \frac{\delta\nu}{\nu}=4\times 10^{-17}\,.
\eeq
Alternatively, one may consider a setup where the distance to the source mass is slowly varied, and the resulting shift in the transition frequency is measured. With the same geometrical parameters as above, only about 1\% of the nuclei are influenced by the field, which suppresses the peak shift averaged over the whole crystal by the same factor. If the averaged peak position can be monitored with a precision of $\delta\nu/\nu\sim 10^{-19}$~\cite{Kazakov_2012}, this approach would achieve a  sensitivity comparable to that of the fast-rotating method based on sidebands. Furthermore, the fast-rotating method possesses the advantage of pushing the signal to a higher frequency at which slow drifts in clock frequency will not hinder the effect.

Several nuclear physics effects could introduce a frequency shift when the sourcing body is brought close to the crystal. Our setup is closely related to the one discussed in~\cite{Gratta:2020hyp}, which explores the use of M\"ossbauer spectroscopy to detect such a frequency shift.
In that work, various systematic sources of error are analyzed, with the dominant contribution arising from the electric field associated with the Casimir effect. This field can displace the electrons surrounding the thorium nucleus, leading to a field shift. 

The Casimir force originates from vacuum fluctuations of the electromagnetic field under specific boundary conditions~\cite{Casimir:1948dh}. As a boundary-dependent phenomenon, it is primarily influenced by the surface electronic properties of the materials used.
Its impact can be mitigated using the iso-electronic technique~\cite{Decca:2005qz}, which involves comparing two dissimilar source masses coated with the same surface material. By applying a uniform gold coating of thickness  $135\,$nm, the differential Casimir force between gold and germanium can be suppressed by six orders of magnitude~\cite{Decca:2005qz,PhysRevA.64.042102}. This suppression technique can be seamlessly incorporated into our proposed experiment. Instead of using a perforated disk, one could employ a disk composed of alternating segments made of materials with significantly different densities e.g. gold and silicon (to maximize the difference in $\delta\phi$), while covering the entire surface  with a thin layer of gold~\cite{Chen:2014oda}. This approach would effectively reduce the influence of the Casimir effect while preserving the desired modulation of the nuclear clock frenquency.

%
\subsection{Host crystals that differ by isotope}
\label{sec:host crystals that differ by isotope}
%
For scalar masses larger than $m_\phi\sim 1\,$eV, traditional fifth-force bounds weaken significantly, opening up opportunities to probe smaller couplings with novel approaches. A natural lower bound on the interaction range is set by the lattice spacing of nuclei in crystals, which typically falls within the $0.1$--$1\,$nm range. Recent advancements in the study of isomeric transition in $^{229}$Th have leveraged solid-state environments, embedding thorium nuclei into specific host crystals such as  LiSrAlF$_6$~\cite{Elwell:2024qyh} and CaF$_2$~\cite{Tiedau:2024obk}, with other potential candidates under consideration~\cite{Hehlen2013OpticalSO}. This technique enables thorium doping densities as high as $10^{20}$ nuclei$/$cm$^3$, significantly enhancing fluorescence rates and facilitating precise frequency measurements of the isomeric transition\cite{Beeks:2022dnl,Zhang:2024dqw}. It also presents a unique opportunity to investigate new forces at sub-nm length scales. 

In such a setup, the crystal-embedded thorium nucleus experiences a field $\phi$ sourced by the surrounding nuclei within the host material. This allows for a comparison of the isomeric transition frequency in hosts of different material densities. One possible approach involves comparing the frequency difference between CaF$_2$ and SrF$_2$ crystals. These materials share similar structures, but since strontium is roughly twice as heavy as calcium, the resulting contrast maximizes  $\delta\phi$. However, disparities in electric field gradients and electronic densities, arising from the different nuclear charges, could introduce deviations at the $\mathcal{O}(1)$ level in the quadrupole splitting and field shift, which are $\lesssim1\,$GHz~\cite{Dessovic:2014,Dzuba:2023muh}. This limitation constrains effective sensitivities to approximately $\delta\nu/\nu\sim 10^{-6}$.

A more refined approach involves comparing two CaF$_2$ crystals that differ only in their calcium isotopes. Specifically, crystals primarily made form $^{40}$Ca or $^{48}$Ca provide an ideal test case, as both calcium isotopes are doubly magic, and their charge radii differ by just $\sim10^{-4}$~\cite{PhysRevLett.115.053003}. This minimizes the nuclear-induced systematic effects, though effects stemming from the mass difference must still be considered.

The primary contributor to $\delta\phi$ in this case is the difference in isotope mass. In thorium-doped CaF$_2$ crystals, Th$^{4+}$ ions replace Ca$^{2+}$ ions, which occupy body-centered cubic lattice sites with a lattice constant of approximately $a\approx 0.57\,$nm~\cite{Gong.et.al}. For scalar masses with Compton wavelengths much larger than the lattice spacing, the thorium nucleus interacts with a large number of surrounding atoms. As a result, the material can be approximated as a homogeneous medium of density $\rho$, with the thorium nucleus positioned at the center of a spherical cavity of radius $a$. Using the field solutions from Sec.~\cref{sec:field solution from a macroscopic probe}, the scalar field influencing the thorium nucleus can be expressed as,
\begin{equation}
    \phi=-Q\cdot d\frac{\rho}{\Mpl m_{\phi}^2}\left(1+m_\phi a\right)e^{-m_\phi a}\,.
\end{equation}
For $m_\phi a\gtrsim1$, deviations from this approximation may arise due to the discrete and inhomogeneous distribution of neighboring isotopes, which depends on factors such as charge compensation mechanisms for the thorium dopant~\cite{Dessovic:2014}. Nevertheless, the exponential suppression remains the dominant effect at higher masses $m_\phi a\gtrsim1$, justifying the use of this result across the relevant mass range. 
A measurement of the isomer frequency difference between $^{48}$CaF$_2$ and $^{40}$CaF$_2$ hosts would be sensitive to,
\beq
|d_g|\approx 7\times 10^{12}\, \left[\frac{8}{\Delta A}\right]^{1/2}\left[\frac{m_\phi}{100\,{\rm eV}}\right]\left[\frac{\delta\nu/\nu}{10^{-10}}\right]^{1/2}\,,
\eeq
for scalar masses $m_\phi\lesssim 200\,$eV.  In deriving this estimate, we have assumed a frequency sensitivity limited by indistinguishable nuclear effects at the level of $\delta\nu/\nu\sim 10^{-10}$ (see discussion below). Due to this limitation the expected reach stays two orders of magnitude behind the 
 bounds from neutron scattering experiments~\cite{Nesvizhevsky:2007by} when considering the coupling $d_g$ and one order of magnitude for $d_e$.

As previously discussed, differences in the charge radius between calcium isotopes may lead to variations in the field shift if they affect the electron density at the thorium nucleus. Additionally, isotope composition directly influences the mechanical properties of the crystal, including the lattice spacing and the velocity at which the emitting nuclei oscillate around the minimum of the lattice potential. These effects are expected to become a dominant systematic background to the new physics signal described above. 

One such background effect is the second-order Doppler shift, which arises from changes in the velocity of the emitting nuclei. In a Debye solid made from a single element where the emitter has a mass difference $\delta m$ relative to the bulk nuclei (which have mass $m$), the frequency shift is given by~\cite{deWaard:1970,Lipkin:1963},
\beq
    \frac{\delta \nu}{\nu} = 3.4\times 10^{-3}\frac{\delta m}{m}\frac{\theta_{\rm D}}{m}\,,
\eeq
where $\theta_{\rm D}$ is the Debye temperature. While the case of CaF$_2$ doped with thorium is more complex, since the bulk consists of different atomic species and thorium is an impurity in both $^{40}$CaF$_2$ and $^{48}$CaF$_2$, this formula provides a rough estimate of the magnitude of the effect. Using $\theta_{\rm D}\simeq\SI{510}{\K}$~\cite{Huffman:1960}, $\delta m/m\simeq0.2$ for the relative mass difference between $^{40}$Ca and $^{48}$Ca, and setting $m$ to the thorium mass, we estimate a shift of $\delta \nu/\nu\simeq 1.4\times 10^{-16}$.

The isotope composition of crystals also influences the lattice constant $a$. 
According to~\cite{VSKogan_1963}, for materials with similar binding force, the dependence of the unit cell volume $V$ (or equivalently, the lattice constant $a$) on the reduced mass $\mu$ follows an approximate linear relation,
\beq
    \frac{\delta V}{V}=3\frac{\delta a}{a}\simeq C \frac{\delta \mu}{\mu}\,,
\eeq
where $C=3\times 10^{-2}$ for ionic crystals and $C=8\times 10^{-3}$ for metals. Estimating $\delta \mu/\mu\simeq  \delta m/m \approx0.2$, we obtain $\delta a/a\approx 2\times10^{-3}$.

A change in the lattice constant affects the quadrupole splitting. The electric field gradient responsible for this splitting originates from the two fluoride ions that compensate for the additional charge of Th$^{4+}$ replacing a Ca$^{2+}$, or alternatively from a calcium vacancy~\cite{Dessovic:2014}. Modeling this gradient as originating from a single point charge at a distance proportional to the lattice constant $a$, we obtain the relation,
\beq
\frac{\delta V_{zz}}{V_{zz}}=-3\frac{\delta a}{a}\,,
\eeq
where $V_{zz}$ is the second derivative of the electromagnetic potential along the $z$-axis. Given that quadrupole splitting is proportional to $V_{zz}$ and is of order $\mathcal{O}(\SI{100}{MHz})$~\cite{Dessovic:2014,Zhang:2024ngu}, the expected shift is $\mathcal{O}(\SI{1}{MHz})$, corresponding to $\delta\nu/\nu=\mathcal{O}(10^{-9}-10^{-10})$. However, by averaging over the multiple transition frequencies resulting from the quadrupole splitting, it is possible to correct for this effect~\cite{Zhang:2024ngu}. Alternatively, one could consider a highly symmetric crystal site, ensuring that the thorium nucleus experiences no gradient of the electric field. The doping sites in CaF$_2$ have such a high symmetry,  if the additional charge of the Th$^{4+}$ is not compensated locally \cite{PhysRevB.111.115103}.  The resulting defect structure has been observed through optical absorption \cite{Beeks:2023mau} and, more directly, laser excitation of $^{229}$Th in CaF$_2$ crystals revealed the absence of the quadrupole splitting at some doping sites \cite{Hiraki:2025uur}. 

Changes in the lattice constant also modify the electron density at the thorium nucleus, affecting the energy required for nuclear charge rearrangement and thereby inducing a field shift between isotopically distinct crystals. 
This shift is primarily driven by $s$-electrons, which have the largest nuclear overlap. 
The valence electron distribution, which is most sensitive to lattice spacing, is expected to be the dominant contributor~\cite{Shirley:1964}. 
To estimate the size of this effect, we note that a change in the ionization state of thorium induces  a shift of $\delta\nu/\nu=\mathcal{O}(10^{-7})$~\cite{Dzuba:2023muh}. 
Ionization corresponds to a complete removal of the valence electrons, leading to a relative density change of ${\cal O}(1)$. 
Assuming that, in a crystal, the valence electron density scales as $1/V$, a change in lattice constant would suppress the effect by $\delta V/V$, yielding an estimated shift of $\delta\nu/\nu=\mathcal{O}(10^{-9}-10^{-10})$.
In~\cref{sec:Moessbauer classic}, we apply a similar estimate to the \SI{93}{\keV} isomer transition in $^{67}$Zn and find that the predicted shift exceeds experimental limits by an order of magnitude. This suggests that our rough estimate may serve as an upper bound on the actual shift. Notably, in both ZnO and Th:CaF$_{2}$, the outer $s$-electrons of the isomeric atom are bound to anions. Since these outer $s$-electrons strongly overlap with the nuclear core and are highly environment-sensitive, this may explain the discrepancy between our estimate and empirical observations.
\newpage
While these systematic effects may limit the sensitivity of this method, they also present interesting avenues for further study. The field shift and quadrupole splitting primarily affect the electric component of the isomer transition energy. 
Their relative significance is enhanced due to the cancellation between electric and nuclear contributions, similar to the mechanism that enhances sensitivity to new physics.
To date, isotope-induced frequency shifts in solid-state environments have not been observed in conventional M\"ossbauer spectroscopy\footnote{Other properties related to M\"ossbauer spectroscopy like the recoil-free fraction have been observed to change due to substitution of isotopes \cite{PhysRevB.60.3005}.}. Investigating such effects could therefore be an interesting scientific target in its own right. Given that the absolute frequency shifts in both quadrupole splitting and field shift could reach $\delta\nu = \mathcal{O}(0.1-\SI{1}{MHz})$, they may already be detectable with existing experimental setups~\cite{Zhang:2024ngu}. Further note that besides the shift of the mean frequency when changing the isotope composition of the crystal, the effects discussed above may lead to a broadening of the linewidth in an isotopically non-pure crystal. In this case the isotope distribution varies around each thorium site inducing local frequency offsets. We would, however, expect the effect of a rearrangement of isotopes to be much smaller than replacing them. An estimate of the induced width is left for future work.

Finally, we note that a more detailed investigation of the systematic errors and technical challenges associated with the two methods discussed for short-distance searches may reveal that a combination of both yields optimal results. For instance, laser excitation of the thorium isomer transition has recently been demonstrated in ThF$_4$ films with thicknesses of $\mathcal{O}(\SI{10}{\nm})$ on various substrates~\cite{Zhang:2024dqw}.
This setup could serve as a static alternative to the spinning disk, enabling searches at comparable distances, as differences in the substrate would induce an EP-violating shift. Note however that substrate induced strain may broaden the line significantly and so far narrow comb spectroscopy in these films is absent.

\section{Conclusions}
\label{sec:conclusion}
We demonstrate that nuclear clocks, particularly solid-state variants, can serve as quintessometers, namely quantum sensors with unprecedented sensitivity to fifth-forces. 
We show how nuclear clocks can probe new
scalar force mediators at submicron scales, unveiling uncharted regions of coupling and mass, while also significantly enhancing equivalence-principle tests at kilometer scales and beyond. Furthermore, we highlight the potential of transportable nuclear clocks to detect scalar interactions over distances exceeding $10\,$km, complementing space-based missions.
Finally, we discuss how nuclear effects in lattices with different isotopes can be detected, leveraging the high level of degeneracy between the excited isomeric state of $^{229}$Th and its ground state.

\section*{Acknowledgement}

We thank Seoul Particle Theory Workshop 2024 for inspiring this work. SJL and BY are supported by the Samsung Science Technology Foundation under Project Number SSTF-BA2201-06. BY is also supported by the NSF grant PHY-2309456. The work of C.\,D. is supported by the CNRS IRP NewSpec. 
The work of GP is supported by grants from NSF-BSF, ISF and Minerva.

\appendix
\section{M\"ossbauer spectroscopy}
\label{sec:Moessbauer classic}
In this section, we derive a bound from the results of~\cite{Potzel:1988}. The experimental setup discussed in~\cref{sec:host crystals that differ by isotope} involves a laser locked to the isomer transition of thorium in a host crystal. The laser frequency is then compared to that of the isomer transition in a crystal with a different  isotope composition (the M\"ossbauer absorber). Here, however, we consider conventional M\"ossbauer spectroscopy, where the light source is not a locked laser but rather spontaneous emission from nuclei in a crystal (the M\"ossbauer emitter). Similar setups have been proposed for new physics searches in~\cite{Gratta:2020hyp,Banerjee:2024bkp}.

Ref.~\cite{Potzel:1988} reports high-resolution M\"ossbauer spectroscopy measurements using a ZnO mono-crystal as the emitter and ZnO powder as the absorber. To enhance the count rate, the  concentration of $^{67}$Zn in the powder was increased from its natural abundance of 4.0\% to 85.2\%. The most abundant Zn isotopes are $^{64}$Zn (49.2\%), $^{66}$Zn (27.7\%) and $^{68}$Zn (18.5\%). Even under the conservative assumption that the relative abundance of these isotopes remains unchanged in the powder, the absorber material is still  slightly heavier (by approximately $1.5\%$), potentially inducing frequency shifts described in~\cref{sec:host crystals that differ by isotope}. 

The highest-resolution scans revealed a relative frequency shift of $(92\pm1)\times 10^{-18}$.
Similar shifts, approximately half this magnitude, were observed in~\cite{Katila:1980} when comparing  a monocrystal and a non-enriched powder. Since the powder in~\cite{Katila:1980} was obtained by grinding a monocrystal that exhibited no significant shift when compared to the emitting monocrystal,  one may argue that all observed shifts are likely due to deformations caused by the grinding process rather than isotope enrichment. The authors of~\cite{Katila:1980} therefore suggest that a M\"ossbauer experiment comparing two monocrystals, one of which is enriched, is necessary to conclusively measure the shift due to isotope effects alone. To establish a bound,  we conservatively assume that shifts larger than $\delta\nu/\nu=10^{-16}$ are excluded. Using the sensitivity coefficients for Zn from~\cite{Fadeev:2021odr}, $K_e=-0.37$, and assuming $K_g\sim 1$, we present the excluded region in~\cref{fig:small_dist} in light blue. However, this bound remains weaker than existing constraints from neutron scattering.

As with the thorium-doped CaF$_2$, it is useful to examine the size of nuclear effects. The second-order Doppler shift due to isotope changes in ZnO was estimated in~\cite{deWaard:1970} to be of order $\delta\nu/\nu\sim2\times10^{-17}$. 

The quadrupole splitting was measured in~\cite{Potzel:1988} to be approximately $\Delta\nu/\nu\sim10^{-13}$. Following the relation from~\cite{VSKogan_1963}, the volume change of the unit cell due to density variation is estimated as $\delta V/V\sim 5\times 10^{-4}$. Assuming the quadrupole splitting scales as $\Delta\nu\propto 1/V$, we estimate a frequency shift of  $\delta\nu/\nu\sim5\times10^{-17}$ between the enriched and non-enriched samples.

The field-shift effect due to lattice constant variations can be estimated as~\cite{Higgins:2024fir,Shirley:1964}
\beq
    \frac{\delta\nu}{\nu}=\frac{(4\pi)^2 Z\alpha}{10\nu}S'(Z)\delta\psi^2(0)\delta R^2\,,
\eeq
where $Z=30$ is the nuclear charge number, $\alpha\approx 1/137$ is the fine-structure constant, and $S'(Z)\approx1.4$ is a relativistic correction factor from~\cite{Shirley:1964}. The charge radius change in the isomer transition is $\delta R^2\approx\SI{0.018}{fm^2}$~\cite{Fadeev:2021odr}. The electron density at the nucleus is estimated as $\psi^2(0)\sim\SI{2e31}{m^{-3}}$ for nuclei with similar charge $Z$~\cite{Shirley:1964}, and we assume it varies as $\delta\psi^2(0)/\psi^2(0)\sim-\delta V/V$. This leads to an estimated shift of $\delta\nu/\nu\sim2\times10^{-15}$. Notably, this  estimate predicts a shift about an order of magnitude larger than the observed value. This discrepancy suggests that more detailed modeling is needed  to accurately predict the effect's size.

\bibliographystyle{utphys28mod}
\bibliography{nukebib}

\providecommand{\href}[2]{#2}\begingroup\raggedright\begin{thebibliography}{10}

\bibitem{Thirolf:2024xlx}
P.~G.~Thirolf, S.~Kraemer, D.~Moritz, and K.~Scharl, ``{The thorium isomer $^{229m}$Th: review of status and perspectives after more than 50 years of research},'' \href{https://dx.doi.org/10.1140/epjs/s11734-024-01098-2}{Eur.\  Phys.\  J.\  ST {\bfseries 233} (2024) 1113--1131}.

\bibitem{Seiferle:2019fbe}
B.~Seiferle {\em et~al.}, ``{Energy of the$^{229}$Th nuclear clock transition},'' \href{https://dx.doi.org/10.1038/s41586-019-1533-4}{Nature {\bfseries 573} (2019) 243--246} {\ttfamily [\href{https://arxiv.org/abs/1905.06308}{arXiv:1905.06308}]}.

\bibitem{Campbell:2012zzb}
C.~J.~Campbell, {\em et al.}, ``{A Single-Ion Nuclear Clock for Metrology at the 19th Decimal Place},'' \href{https://dx.doi.org/10.1103/PhysRevLett.108.120802}{Phys.\  Rev.\  Lett.\  {\bfseries 108} (2012) 120802} {\ttfamily [\href{https://arxiv.org/abs/1110.2490}{arXiv:1110.2490}]}.

\bibitem{Kazakov_2012}
G.~A.~Kazakov, {\em et al.}, ``Performance of a 229Thorium solid-state nuclear clock,'' \href{https://dx.doi.org/10.1088/1367-2630/14/8/083019}{New Journal of Physics {\bfseries 14} (2012) 083019} {\ttfamily [\href{https://arxiv.org/abs/1204.3268}{arXiv:1204.3268}]}, \url{https://dx.doi.org/10.1088/1367-2630/14/8/083019}.

\bibitem{Helmer:1994zz}
R.~G.~Helmer and C.~W.~Reich, ``{An excited state of Th-229 at 3.5 eV},'' \href{https://dx.doi.org/10.1103/PhysRevC.49.1845}{Phys.\  Rev.\  C {\bfseries 49} (1994) 1845--1858}.

\bibitem{Matinyan:1997ih}
S.~G.~Matinyan, ``{Lasers as a bridge between atomic and nuclear physics},'' \href{https://dx.doi.org/10.1016/S0370-1573(97)00084-7}{Phys.\  Rept.\  {\bfseries 298} (1998) 199} {\ttfamily [\href{https://arxiv.org/abs/nucl-th/9706005}{nucl-th/9706005}]}.

\bibitem{Sikorsky:2020peq}
T.~Sikorsky {\em et~al.}, ``{Measurement of the $^{229}$Th isomer energy with a magnetic micro-calorimeter},'' \href{https://dx.doi.org/10.1103/PhysRevLett.125.142503}{Phys.\  Rev.\  Lett.\  {\bfseries 125} (2020) 142503} {\ttfamily [\href{https://arxiv.org/abs/2005.13340}{arXiv:2005.13340}]}.

\bibitem{Kraemer:2022gpi}
S.~Kraemer {\em et~al.}, ``{Observation of the radiative decay of the $^{229}$Th nuclear clock isomer},'' \href{https://dx.doi.org/10.1038/s41586-023-05894-z}{Nature {\bfseries 617} (2023) 706--710} {\ttfamily [\href{https://arxiv.org/abs/2209.10276}{arXiv:2209.10276}]}.

\bibitem{Tiedau:2024obk}
J.~Tiedau {\em et~al.}, ``{Laser Excitation of the Th-229 Nucleus},'' \href{https://dx.doi.org/10.1103/PhysRevLett.132.182501}{Phys.\  Rev.\  Lett.\  {\bfseries 132} (2024) 182501}.

\bibitem{Elwell:2024qyh}
R.~Elwell, {\em et al.}, ``{Laser Excitation of the Th229 Nuclear Isomeric Transition in a Solid-State Host},'' \href{https://dx.doi.org/10.1103/PhysRevLett.133.013201}{Phys.\  Rev.\  Lett.\  {\bfseries 133} (2024) 013201} {\ttfamily [\href{https://arxiv.org/abs/2404.12311}{arXiv:2404.12311}]}.

\bibitem{Zhang:2024ngu}
C.~Zhang {\em et~al.}, ``{Frequency ratio of the $^{229m}$Th nuclear isomeric transition and the $^{87}$Sr atomic clock},'' \href{https://dx.doi.org/10.1038/s41586-024-07839-6}{Nature {\bfseries 633} (2024) 63--70} {\ttfamily [\href{https://arxiv.org/abs/2406.18719}{arXiv:2406.18719}]}.

\bibitem{Flambaum:2006ak}
V.~V.~Flambaum, ``{Enhanced effect of temporal variation of the fine structure constant and the strong interaction in Th-229},'' \href{https://dx.doi.org/10.1103/PhysRevLett.97.092502}{Phys.\  Rev.\  Lett.\  {\bfseries 97} (2006) 092502} {\ttfamily [\href{https://arxiv.org/abs/physics/0601034}{physics/0601034}]}.

\bibitem{Fadeev:2020bgt}
P.~Fadeev, J.~C.~Berengut, and V.~V.~Flambaum, ``{Sensitivity of $^{229}$Th nuclear clock transition to variation of the fine-structure constant},'' \href{https://dx.doi.org/10.1103/PhysRevA.102.052833}{Phys.\  Rev.\  A {\bfseries 102} (2020) 052833} {\ttfamily [\href{https://arxiv.org/abs/2007.00408}{arXiv:2007.00408}]}.

\bibitem{Minkov:2021ovq}
N.~Minkov and A.~P{\'a}lffy, ``{$^{229}$Thm isomer from a nuclear model perspective},'' \href{https://dx.doi.org/10.1103/PhysRevC.103.014313}{Phys.\  Rev.\  C {\bfseries 103} (2021) 014313} {\ttfamily [\href{https://arxiv.org/abs/2102.02288}{arXiv:2102.02288}]}.

\bibitem{Fadeev:2021odr}
P.~Fadeev, J.~C.~Berengut, and V.~V.~Flambaum, ``{Effects of variation of the fine structure constant \ensuremath{\alpha} and quark mass mq in M\"ossbauer nuclear transitions},'' \href{https://dx.doi.org/10.1103/PhysRevC.105.L051303}{Phys.\  Rev.\  C {\bfseries 105} (2022) L051303} {\ttfamily [\href{https://arxiv.org/abs/2112.13195}{arXiv:2112.13195}]}.

\bibitem{Caputo:2024doz}
A.~Caputo, {\em et al.}, ``{Sensitivity of nuclear clocks to new physics},'' \href{https://dx.doi.org/10.1103/l29n-gt5j}{Phys.\  Rev.\  C {\bfseries 112} (2025) L031302} {\ttfamily [\href{https://arxiv.org/abs/2407.17526}{arXiv:2407.17526}]}.

\bibitem{Beeks:2024xnc}
K.~Beeks {\em et~al.}, ``{Fine-structure constant sensitivity of the Th-229 nuclear clock transition}.'' {\ttfamily \href{https://arxiv.org/abs/2407.17300}{arXiv:2407.17300}}.

\bibitem{Minkov:2024wna}
N.~Minkov, A.~P{\'a}lffy, P.~Quentin, and L.~Bonneau, ``{Skyrme-Hartree-Fock-BCS approach to Th229m and neighboring nuclei},'' \href{https://dx.doi.org/10.1103/PhysRevC.110.034327}{Phys.\  Rev.\  C {\bfseries 110} (2024) 034327} {\ttfamily [\href{https://arxiv.org/abs/2408.11010}{arXiv:2408.11010}]}.

\bibitem{Arvanitaki:2014faa}
A.~Arvanitaki, J.~Huang, and K.~Van~Tilburg, ``{Searching for dilaton dark matter with atomic clocks},'' \href{https://dx.doi.org/10.1103/PhysRevD.91.015015}{Phys.\  Rev.\  D {\bfseries 91} (2015) 015015} {\ttfamily [\href{https://arxiv.org/abs/1405.2925}{arXiv:1405.2925}]}.

\bibitem{Stadnik:2015kia}
Y.~V.~Stadnik and V.~V.~Flambaum, ``{Can dark matter induce cosmological evolution of the fundamental constants of Nature?}'' \href{https://dx.doi.org/10.1103/PhysRevLett.115.201301}{Phys.\  Rev.\  Lett.\  {\bfseries 115} (2015) 201301} {\ttfamily [\href{https://arxiv.org/abs/1503.08540}{arXiv:1503.08540}]}.

\bibitem{Safronova:2017xyt}
M.~S.~Safronova, {\em et al.}, ``{Search for New Physics with Atoms and Molecules},'' \href{https://dx.doi.org/10.1103/RevModPhys.90.025008}{Rev.\  Mod.\  Phys.\  {\bfseries 90} (2018) 025008} {\ttfamily [\href{https://arxiv.org/abs/1710.01833}{arXiv:1710.01833}]}.

\bibitem{Fuchs:2024xvc}
E.~Fuchs, {\em et al.}, ``{Searching for Dark Matter with the Th229 Nuclear Lineshape from Laser Spectroscopy},'' \href{https://dx.doi.org/10.1103/PhysRevX.15.021055}{Phys.\  Rev.\  X {\bfseries 15} (2025) 021055} {\ttfamily [\href{https://arxiv.org/abs/2407.15924}{arXiv:2407.15924}]}.

\bibitem{Kim:2022ype}
H.~Kim and G.~Perez, ``{Oscillations of atomic energy levels induced by QCD axion dark matter},'' \href{https://dx.doi.org/10.1103/PhysRevD.109.015005}{Phys.\  Rev.\  D {\bfseries 109} (2024) 015005} {\ttfamily [\href{https://arxiv.org/abs/2205.12988}{arXiv:2205.12988}]}.

\bibitem{Banerjee:2019epw}
A.~Banerjee, D.~Budker, J.~Eby, H.~Kim, and G.~Perez, ``{Relaxion Stars and their detection via Atomic Physics},'' \href{https://dx.doi.org/10.1038/s42005-019-0260-3}{Commun.\  Phys.\  {\bfseries 3} (2020) 1} {\ttfamily [\href{https://arxiv.org/abs/1902.08212}{arXiv:1902.08212}]}.

\bibitem{Banerjee:2019xuy}
A.~Banerjee, {\em et al.}, ``{Searching for Earth/Solar Axion Halos},'' \href{https://dx.doi.org/10.1007/JHEP09(2020)004}{JHEP {\bfseries 09} (2020) 004} {\ttfamily [\href{https://arxiv.org/abs/1912.04295}{arXiv:1912.04295}]}.

\bibitem{Budker:2023sex}
D.~Budker, J.~Eby, M.~Gorghetto, M.~Jiang, and G.~Perez, ``{A generic formation mechanism of ultralight dark matter solar halos},'' \href{https://dx.doi.org/10.1088/1475-7516/2023/12/021}{JCAP {\bfseries 12} (2023) 021} {\ttfamily [\href{https://arxiv.org/abs/2306.12477}{arXiv:2306.12477}]}.

\bibitem{Derevianko:2013oaa}
A.~Derevianko and M.~Pospelov, ``{Hunting for topological dark matter with atomic clocks},'' \href{https://dx.doi.org/10.1038/nphys3137}{Nature Phys.\  {\bfseries 10} (2014) 933} {\ttfamily [\href{https://arxiv.org/abs/1311.1244}{arXiv:1311.1244}]}.

\bibitem{Damour:1999ip}
T.~Damour, ``{Equivalence principle and clocks},'' in {\em {34th Rencontres de Moriond: Gravitational Waves and Experimental Gravity}}, pp.~357--365.
\newblock 1999.
\newblock {\ttfamily \href{https://arxiv.org/abs/gr-qc/9904032}{gr-qc/9904032}}.

\bibitem{Damour:2010rp}
T.~Damour and J.~F.~Donoghue, ``{Equivalence Principle Violations and Couplings of a Light Dilaton},'' \href{https://dx.doi.org/10.1103/PhysRevD.82.084033}{Phys.\  Rev.\  D {\bfseries 82} (2010) 084033} {\ttfamily [\href{https://arxiv.org/abs/1007.2792}{arXiv:1007.2792}]}.

\bibitem{Oswald:2021vtc}
R.~Oswald {\em et~al.}, ``{Search for Dark-Matter-Induced Oscillations of Fundamental Constants Using Molecular Spectroscopy},'' \href{https://dx.doi.org/10.1103/PhysRevLett.129.031302}{Phys.\  Rev.\  Lett.\  {\bfseries 129} (2022) 031302} {\ttfamily [\href{https://arxiv.org/abs/2111.06883}{arXiv:2111.06883}]}.

\bibitem{Banerjee:2022sqg}
A.~Banerjee, G.~Perez, M.~Safronova, I.~Savoray, and A.~Shalit, ``{The phenomenology of quadratically coupled ultra light dark matter},'' \href{https://dx.doi.org/10.1007/JHEP10(2023)042}{JHEP {\bfseries 10} (2023) 042} {\ttfamily [\href{https://arxiv.org/abs/2211.05174}{arXiv:2211.05174}]}.

\bibitem{aristotle_heavens}
Aristotle, {\em On the Heavens}.
\newblock circa 350 BCE.
\newblock Estimated date of writing.

\bibitem{Spero:1980zz}
R.~Spero, J.~K.~Hoskins, R.~Newman, J.~Pellam, and J.~Schultz, ``{Test of the Gravitational Inverse-Square Law at Laboratory Distances},'' \href{https://dx.doi.org/10.1103/PhysRevLett.44.1645}{Phys.\  Rev.\  Lett.\  {\bfseries 44} (1980) 1645--1648}.

\bibitem{Mostepanenko:1990ed}
V.~M.~Mostepanenko and I.~Y.~Sokolov, ``{Stronger restrictions on the constants of long range forces decreasing as r**(-n)},'' \href{https://dx.doi.org/10.1016/0375-9601(90)90715-Z}{Phys.\  Lett.\  A {\bfseries 146} (1990) 373--374}.

\bibitem{Bordag:1993htq}
M.~Bordag, V.~M.~Mostepanenko, and I.~Y.~Sokolov, ``{On the strengthening of restrictions on hypothetical Yukawa type forces with extremely small range of action},'' \href{https://dx.doi.org/10.1016/0375-9601(94)90860-5}{Phys.\  Lett.\  A {\bfseries 187} (1994) 35--39}.

\bibitem{Lamoreaux:1996wh}
S.~K.~Lamoreaux, ``{Demonstration of the Casimir force in the 0.6 to 6 micrometers range},'' \href{https://dx.doi.org/10.1103/PhysRevLett.78.5}{Phys.\  Rev.\  Lett.\  {\bfseries 78} (1997) 5--8}. [Erratum: Phys.Rev.Lett. 81, 5475--5476 (1998)].

\bibitem{Bordag:1998nv}
M.~Bordag, B.~Geyer, G.~L.~Klimchitskaya, and V.~M.~Mostepanenko, ``{Constraints for hypothetical interactions from a recent demonstration of the Casimir force and some possible improvements},'' \href{https://dx.doi.org/10.1103/PhysRevD.58.075003}{Phys.\  Rev.\  D {\bfseries 58} (1998) 075003} {\ttfamily [\href{https://arxiv.org/abs/hep-ph/9804223}{hep-ph/9804223}]}.

\bibitem{Ederth:2000zz}
T.~Ederth, ``{Template-stripped gold surfaces with 0.4-nm rms roughness suitable for force measurements: Application to the Casimir force in the 20-100-nm range},'' \href{https://dx.doi.org/10.1103/PhysRevA.62.062104}{Phys.\  Rev.\  A {\bfseries 62} (2000) 062104} {\ttfamily [\href{https://arxiv.org/abs/quant-ph/0008009}{quant-ph/0008009}]}.

\bibitem{Harris:2000zz}
B.~W.~Harris, F.~Chen, and U.~Mohideen, ``{Precision measurement of the Casimir force using gold surfaces},'' \href{https://dx.doi.org/10.1103/PhysRevA.62.052109}{Phys.\  Rev.\  A {\bfseries 62} (2000) 052109} {\ttfamily [\href{https://arxiv.org/abs/quant-ph/0005088}{quant-ph/0005088}]}.

\bibitem{Fischbach:2001ry}
E.~Fischbach, D.~E.~Krause, V.~M.~Mostepanenko, and M.~Novello, ``{New constraints on ultrashort ranged Yukawa interactions from atomic force microscopy},'' \href{https://dx.doi.org/10.1103/PhysRevD.64.075010}{Phys.\  Rev.\  D {\bfseries 64} (2001) 075010} {\ttfamily [\href{https://arxiv.org/abs/hep-ph/0106331}{hep-ph/0106331}]}.

\bibitem{Chiaverini:2002cb}
J.~Chiaverini, S.~J.~Smullin, A.~A.~Geraci, D.~M.~Weld, and A.~Kapitulnik, ``{New experimental constraints on nonNewtonian forces below 100 microns},'' \href{https://dx.doi.org/10.1103/PhysRevLett.90.151101}{Phys.\  Rev.\  Lett.\  {\bfseries 90} (2003) 151101} {\ttfamily [\href{https://arxiv.org/abs/hep-ph/0209325}{hep-ph/0209325}]}.

\bibitem{Adelberger:2003zx}
E.~G.~Adelberger, B.~R.~Heckel, and A.~E.~Nelson, ``{Tests of the gravitational inverse square law},'' \href{https://dx.doi.org/10.1146/annurev.nucl.53.041002.110503}{Ann.\  Rev.\  Nucl.\  Part.\  Sci.\  {\bfseries 53} (2003) 77--121} {\ttfamily [\href{https://arxiv.org/abs/hep-ph/0307284}{hep-ph/0307284}]}.

\bibitem{Long:2003dx}
J.~C.~Long, {\em et al.}, ``{Upper limits to submillimeter-range forces from extra space-time dimensions},'' \href{https://dx.doi.org/10.1038/nature01432}{Nature {\bfseries 421} (2003) 922--925} {\ttfamily [\href{https://arxiv.org/abs/hep-ph/0210004}{hep-ph/0210004}]}.

\bibitem{Kapner:2006si}
D.~J.~Kapner, {\em et al.}, ``{Tests of the gravitational inverse-square law below the dark-energy length scale},'' \href{https://dx.doi.org/10.1103/PhysRevLett.98.021101}{Phys.\  Rev.\  Lett.\  {\bfseries 98} (2007) 021101} {\ttfamily [\href{https://arxiv.org/abs/hep-ph/0611184}{hep-ph/0611184}]}.

\bibitem{Yang:2012zzb}
S.-Q.~Yang, {\em et al.}, ``{Test of the Gravitational Inverse Square Law at Millimeter Ranges},'' \href{https://dx.doi.org/10.1103/PhysRevLett.108.081101}{Phys.\  Rev.\  Lett.\  {\bfseries 108} (2012) 081101}.

\bibitem{Chen:2014oda}
Y.~J.~Chen, {\em et al.}, ``{Stronger Limits on Hypothetical Yukawa Interactions in the 30\textendash{}8000 nm Range},'' \href{https://dx.doi.org/10.1103/PhysRevLett.116.221102}{Phys.\  Rev.\  Lett.\  {\bfseries 116} (2016) 221102} {\ttfamily [\href{https://arxiv.org/abs/1410.7267}{arXiv:1410.7267}]}.

\bibitem{Leefer:2016xfu}
N.~Leefer, A.~Gerhardus, D.~Budker, V.~V.~Flambaum, and Y.~V.~Stadnik, ``{Search for the effect of massive bodies on atomic spectra and constraints on Yukawa-type interactions of scalar particles},'' \href{https://dx.doi.org/10.1103/PhysRevLett.117.271601}{Phys.\  Rev.\  Lett.\  {\bfseries 117} (2016) 271601} {\ttfamily [\href{https://arxiv.org/abs/1607.04956}{arXiv:1607.04956}]}.

\bibitem{Tan:2020vpf}
W.-H.~Tan {\em et~al.}, ``{Improvement for Testing the Gravitational Inverse-Square Law at the Submillimeter Range},'' \href{https://dx.doi.org/10.1103/PhysRevLett.124.051301}{Phys.\  Rev.\  Lett.\  {\bfseries 124} (2020) 051301}.

\bibitem{Lee:2020zjt}
J.~G.~Lee, E.~G.~Adelberger, T.~S.~Cook, S.~M.~Fleischer, and B.~R.~Heckel, ``{New Test of the Gravitational $1/r^2$ Law at Separations down to 52 $\mu$m},'' \href{https://dx.doi.org/10.1103/PhysRevLett.124.101101}{Phys.\  Rev.\  Lett.\  {\bfseries 124} (2020) 101101} {\ttfamily [\href{https://arxiv.org/abs/2002.11761}{arXiv:2002.11761}]}.

\bibitem{Brzeminski:2022sde}
D.~Brzeminski, Z.~Chacko, A.~Dev, I.~Flood, and A.~Hook, ``{Searching for a fifth force with atomic and nuclear clocks},'' \href{https://dx.doi.org/10.1103/PhysRevD.106.095031}{Phys.\  Rev.\  D {\bfseries 106} (2022) 095031} {\ttfamily [\href{https://arxiv.org/abs/2207.14310}{arXiv:2207.14310}]}.

\bibitem{Kim:2023pvt}
H.~Kim, A.~Lenoci, G.~Perez, and W.~Ratzinger, ``{Probing an ultralight QCD axion with electromagnetic quadratic interaction},'' \href{https://dx.doi.org/10.1103/PhysRevD.109.015030}{Phys.\  Rev.\  D {\bfseries 109} (2024) 015030} {\ttfamily [\href{https://arxiv.org/abs/2307.14962}{arXiv:2307.14962}]}.

\bibitem{Derevianko:2021kye}
A.~Derevianko, {\em et al.}, ``{Fundamental physics with a state-of-the-art optical clock in space},'' \href{https://dx.doi.org/10.1088/2058-9565/ac7df9}{Quantum Sci.\  Technol.\  {\bfseries 7} (2022) 044002} {\ttfamily [\href{https://arxiv.org/abs/2112.10817}{arXiv:2112.10817}]}.

\bibitem{Schlamminger:2007ht}
S.~Schlamminger, K.~Y.~Choi, T.~A.~Wagner, J.~H.~Gundlach, and E.~G.~Adelberger, ``{Test of the equivalence principle using a rotating torsion balance},'' \href{https://dx.doi.org/10.1103/PhysRevLett.100.041101}{Phys.\  Rev.\  Lett.\  {\bfseries 100} (2008) 041101} {\ttfamily [\href{https://arxiv.org/abs/0712.0607}{arXiv:0712.0607}]}.

\bibitem{Smith:1999cr}
G.~L.~Smith, {\em et al.}, ``{Short range tests of the equivalence principle},'' \href{https://dx.doi.org/10.1103/PhysRevD.61.022001}{Phys.\  Rev.\  D {\bfseries 61} (2000) 022001} {\ttfamily [\href{https://arxiv.org/abs/2405.10982}{arXiv:2405.10982}]}.

\bibitem{Berge:2017ovy}
J.~Berg\'e, {\em et al.}, ``{MICROSCOPE Mission: First Constraints on the Violation of the Weak Equivalence Principle by a Light Scalar Dilaton},'' \href{https://dx.doi.org/10.1103/PhysRevLett.120.141101}{Phys.\  Rev.\  Lett.\  {\bfseries 120} (2018) 141101} {\ttfamily [\href{https://arxiv.org/abs/1712.00483}{arXiv:1712.00483}]}.

\bibitem{MICROSCOPE:2022doy}
{\bfseries MICROSCOPE} Collaboration, ``{MICROSCOPE Mission: Final Results of the Test of the Equivalence Principle},'' \href{https://dx.doi.org/10.1103/PhysRevLett.129.121102}{Phys.\  Rev.\  Lett.\  {\bfseries 129} (2022) 121102} {\ttfamily [\href{https://arxiv.org/abs/2209.15487}{arXiv:2209.15487}]}.

\bibitem{Grotti:2017ftq}
J.~Grotti {\em et~al.}, ``{Geodesy and metrology with a transportable optical clock},'' \href{https://dx.doi.org/10.1038/s41567-017-0042-3}{Nature Phys.\  {\bfseries 14} (2018) 437--441} {\ttfamily [\href{https://arxiv.org/abs/1705.04089}{arXiv:1705.04089}]}.

\bibitem{Bordag:2001qi}
M.~Bordag, U.~Mohideen, and V.~M.~Mostepanenko, ``{New developments in the Casimir effect},'' \href{https://dx.doi.org/10.1016/S0370-1573(01)00015-1}{Phys.\  Rept.\  {\bfseries 353} (2001) 1--205} {\ttfamily [\href{https://arxiv.org/abs/quant-ph/0106045}{quant-ph/0106045}]}.

\bibitem{Ooi:2025hmd}
T.~Ooi, {\em et al.}, ``{Frequency reproducibility of solid-state Th-229 nuclear clocks}.'' {\ttfamily \href{https://arxiv.org/abs/2507.01180}{arXiv:2507.01180}}.

\bibitem{Gratta:2020hyp}
G.~Gratta, D.~E.~Kaplan, and S.~Rajendran, ``{Searching for New Interactions at Sub-micron Scale Using the Mossbauer Effect},'' \href{https://dx.doi.org/10.1103/PhysRevD.102.115031}{Phys.\  Rev.\  D {\bfseries 102} (2020) 115031} {\ttfamily [\href{https://arxiv.org/abs/2010.03588}{arXiv:2010.03588}]}.

\bibitem{Casimir:1948dh}
H.~B.~G.~Casimir, ``{On the attraction between two perfectly conducting plates},'' Indag.\  Math.\  {\bfseries 10} (1948) 261--263.

\bibitem{Decca:2005qz}
R.~S.~Decca, {\em et al.}, ``{Constraining new forces in the Casimir regime using the isoelectronic technique},'' \href{https://dx.doi.org/10.1103/PhysRevLett.94.240401}{Phys.\  Rev.\  Lett.\  {\bfseries 94} (2005) 240401} {\ttfamily [\href{https://arxiv.org/abs/hep-ph/0502025}{hep-ph/0502025}]}.

\bibitem{PhysRevA.64.042102}
R.~Matloob and H.~Falinejad, ``Casimir force between two dielectric slabs,'' \href{https://dx.doi.org/10.1103/PhysRevA.64.042102}{Phys.\  Rev.\  A {\bfseries 64} (2001) 042102}, \url{https://link.aps.org/doi/10.1103/PhysRevA.64.042102}.

\bibitem{Hehlen2013OpticalSO}
M.~P.~Hehlen, {\em et al.}, ``Optical spectroscopy of an atomic nucleus: Progress toward direct observation of the 229Th isomer transition,'' Journal of Luminescence {\bfseries 133} (2013) 91--95, \url{https://api.semanticscholar.org/CorpusID:98268442}.

\bibitem{Beeks:2022dnl}
K.~Beeks {\em et~al.}, ``{Growth and characterization of thorium-doped calcium fluoride single crystals},'' \href{https://dx.doi.org/10.1038/s41598-023-31045-5}{Sci.\  Rep.\  {\bfseries 13} (2023) 3897} {\ttfamily [\href{https://arxiv.org/abs/2211.05445}{arXiv:2211.05445}]}.

\bibitem{Zhang:2024dqw}
C.~Zhang {\em et~al.}, ``{$^{229}$ThF$_{4}$ thin films for solid-state nuclear clocks},'' \href{https://dx.doi.org/10.1038/s41586-024-08256-5}{Nature {\bfseries 636} (2024) 603--608} {\ttfamily [\href{https://arxiv.org/abs/2410.01753}{arXiv:2410.01753}]}.

\bibitem{Dessovic:2014}
P.~Dessovic, {\em et al.}, ``229Thorium-doped calcium fluoride for nuclear laser spectroscopy,'' \href{https://dx.doi.org/10.1088/0953-8984/26/10/105402}{Journal of Physics: Condensed Matter {\bfseries 26} (2014) 105402}, \url{https://dx.doi.org/10.1088/0953-8984/26/10/105402}.

\bibitem{Dzuba:2023muh}
V.~A.~Dzuba and V.~V.~Flambaum, ``{Effects of Electrons on Nuclear Clock Transition Frequency in Th229 Ions},'' \href{https://dx.doi.org/10.1103/PhysRevLett.131.263002}{Phys.\  Rev.\  Lett.\  {\bfseries 131} (2023) 263002} {\ttfamily [\href{https://arxiv.org/abs/2309.11176}{arXiv:2309.11176}]}.

\bibitem{PhysRevLett.115.053003}
F.~Gebert, {\em et al.}, ``Precision Isotope Shift Measurements in Calcium Ions Using Quantum Logic Detection Schemes,'' \href{https://dx.doi.org/10.1103/PhysRevLett.115.053003}{Phys.\  Rev.\  Lett.\  {\bfseries 115} (2015) 053003}, \url{https://link.aps.org/doi/10.1103/PhysRevLett.115.053003}.

\bibitem{Gong.et.al}
Q.~Gong, {\em et al.}, ``Structures and Properties of High-Concentration Doped Th:CaF2 Single Crystals for Solid-State Nuclear Clock Materials,'' \href{https://dx.doi.org/10.1021/acs.inorgchem.3c04009}{Inorganic Chemistry {\bfseries 63} (2024) 3807--3814}, \url{https://doi.org/10.1021/acs.inorgchem.3c04009}. PMID: 38345921.

\bibitem{Nesvizhevsky:2007by}
V.~V.~Nesvizhevsky, G.~Pignol, and K.~V.~Protasov, ``{Neutron scattering and extra short range interactions},'' \href{https://dx.doi.org/10.1103/PhysRevD.77.034020}{Phys.\  Rev.\  D {\bfseries 77} (2008) 034020} {\ttfamily [\href{https://arxiv.org/abs/0711.2298}{arXiv:0711.2298}]}.

\bibitem{deWaard:1970}
H.~de~Waard and G.~J.~Perlow, ``M\"ossbauer Effect of the 93-keV Transition in ${\mathrm{Zn}}^{67}$,'' \href{https://dx.doi.org/10.1103/PhysRevLett.24.566}{Phys.\  Rev.\  Lett.\  {\bfseries 24} (1970) 566--569}, \url{https://link.aps.org/doi/10.1103/PhysRevLett.24.566}.

\bibitem{Lipkin:1963}
H.~J.~Lipkin, ``Some simple features of the Mössbauer effect. III: The f-factor for an impurity source in a crystal,'' \href{https://dx.doi.org/https://doi.org/10.1016/0003-4916(63)90251-3}{Annals of Physics {\bfseries 23} (1963) 28--37}, \url{https://www.sciencedirect.com/science/article/pii/0003491663902513}.

\bibitem{Huffman:1960}
D.~R.~Huffman and M.~H.~Norwood, ``Specific Heat and Elastic Constants of Calcium Fluoride at Low Temperatures,'' \href{https://dx.doi.org/10.1103/PhysRev.117.709}{Phys.\  Rev.\  {\bfseries 117} (1960) 709--711}, \url{https://link.aps.org/doi/10.1103/PhysRev.117.709}.

\bibitem{VSKogan_1963}
V.~S.~Kogan, ``ISOTOPE EFFECTS IN THE STRUCTURAL PROPERTIES OF SOLIDS,'' \href{https://dx.doi.org/10.1070/PU1963v005n06ABEH003469}{Soviet Physics Uspekhi {\bfseries 5} (1963) 951}, \url{https://dx.doi.org/10.1070/PU1963v005n06ABEH003469}.

\bibitem{PhysRevB.111.115103}
K.~Nalikowski, V.~Veryazov, K.~Beeks, T.~Schumm, and M.~Kro\ifmmode~\acute{s}\else \'{s}\fi{}nicki, ``Embedded cluster approach for accurate electronic structure calculations of $^{229}\mathrm{Th}\text{:}{\mathrm{CaF}}_{2}$,'' \href{https://dx.doi.org/10.1103/PhysRevB.111.115103}{Phys.\  Rev.\  B {\bfseries 111} (2025) 115103}, \url{https://link.aps.org/doi/10.1103/PhysRevB.111.115103}.

\bibitem{Beeks:2023mau}
K.~Beeks {\em et~al.}, ``{Optical transmission enhancement of ionic crystals via superionic fluoride transfer: Growing VUV-transparent radioactive crystals},'' \href{https://dx.doi.org/10.1103/PhysRevB.109.094111}{Phys.\  Rev.\  B {\bfseries 109} (2024) 094111} {\ttfamily [\href{https://arxiv.org/abs/2312.13713}{arXiv:2312.13713}]}.

\bibitem{Hiraki:2025uur}
T.~Hiraki {\em et~al.}, ``{Laser M{\"o}ssbauer spectroscopy of {\textasciicircum}{229}Th}.'' {\ttfamily \href{https://arxiv.org/abs/2509.00041}{arXiv:2509.00041}}.

\bibitem{Shirley:1964}
D.~A.~SHIRLEY, ``Application and Interpretation of Isomer Shifts,'' \href{https://dx.doi.org/10.1103/RevModPhys.36.339}{Rev.\  Mod.\  Phys.\  {\bfseries 36} (1964) 339--351}, \url{https://link.aps.org/doi/10.1103/RevModPhys.36.339}.

\bibitem{PhysRevB.60.3005}
V.~Chechersky, {\em et al.}, ``Origin of the giant oxygen-isotope effect in the manganite ${\mathrm{La}}_{0.8}{\mathrm{Ca}}_{0.2}{\mathrm{MnO}}_{3}$,'' \href{https://dx.doi.org/10.1103/PhysRevB.60.3005}{Phys.\  Rev.\  B {\bfseries 60} (1999) 3005--3008}, \url{https://link.aps.org/doi/10.1103/PhysRevB.60.3005}.

\bibitem{Potzel:1988}
W.~Potzel, ``{Recent 67Zn-experiments},'' \href{https://dx.doi.org/10.1007/BF02049088}{Hyperfine Interactions {\bfseries 40} (1988) 171–182}, \url{https://doi.org/10.1007/BF02049088}.

\bibitem{Banerjee:2024bkp}
A.~Banerjee, ``{Probing (Ultra-) Light Dark Matter Using Synchrotron Based M{\"o}ssbauer Spectroscopy}.'' {\ttfamily \href{https://arxiv.org/abs/2408.04700}{arXiv:2408.04700}}.

\bibitem{Katila:1980}
{Katila, T. }, {Riski, K.  J. }, and {Ylä-Jääski, J. }, ``PRECISION DETERMINATION OF THE ISOMER SHIFT OF ZN-67 IN ZNO,'' \href{https://dx.doi.org/10.1051/jphyscol:1980125}{J.\  Phys.\  Colloques {\bfseries 41} (1980) C1--121--C1--122}, \url{https://doi.org/10.1051/jphyscol:1980125}.

\bibitem{Higgins:2024fir}
J.~S.~Higgins, {\em et al.}, ``{Temperature Sensitivity of a Thorium-229 Solid-State Nuclear Clock},'' \href{https://dx.doi.org/10.1103/PhysRevLett.134.113801}{Phys.\  Rev.\  Lett.\  {\bfseries 134} (2025) 113801} {\ttfamily [\href{https://arxiv.org/abs/2409.11590}{arXiv:2409.11590}]}.

\end{thebibliography}\endgroup

\end{document}